# Frequency ratio of Yb and Sr clocks with 5×10⁻¹⁷ uncertainty at 150 s averaging time

N. Nemitz[1], T. Ohkubo[1,2,3], M. Takamoto[1,2,4], I. Ushijima[1,2,4], M. Das[1,2,4], N. Ohmae[1,2,3], H. Katori[1,2,3,4*]

[1] Quantum Metrology Laboratory, RIKEN, Wako, Saitama 351-0198, Japan

[2] Innovative Space-Time Project, ERATO, Japan Science and Technology Agency, Bunkyo-ku, Tokyo 113-8656, Japan

[3] Department of Applied Physics, Graduate School of Engineering, The University of Tokyo, Bunkyo-ku, Tokyo 113-8656, Japan

[4] RIKEN Center for Advanced Photonics, Wako, Saitama 351-0198, Japan

* email: katori@amo.t.u-tokyo.ac.jp

**Transition frequencies of atoms and ions are among the most accurately accessible quantities in nature, playing important roles in pushing the frontiers of science by testing fundamental laws of physics, in addition to a wide range of applications such as satellite navigation systems. Atomic clocks based on optical transitions approach uncertainties of $10^{-18}$ (ref. 1-3), where full frequency descriptions are far beyond the reach of the SI second. Frequency ratios of such super clocks, on the other hand, are not subject to this limitation[4-8]. They can therefore verify consistency and overall accuracy for an ensemble of super clocks, an essential step towards a redefinition of the second[9]. However, with the measurement stabilities so far reported for such frequency ratios, a confirmation to $1\times10^{-18}$ uncertainty would require an averaging time $\tau$ of multiple months. Here we report a measurement of the frequency ratio of neutral ytterbium and strontium clocks with a much improved stability of $4\times10^{-16}\,(\tau/\mathrm{s})^{-1/2}$. Enabled by the high stability of optical lattice clocks[10] interrogating hundreds of atoms, this marks a 90-fold reduction in the required averaging time over a previous record-setting experiment[5] that determined the ratio of Al⁺ and Hg⁺ single-ion clocks to an uncertainty of $5.2\times10^{-17}$. For the Yb/Sr ratio, we find $\mathcal{R}$ = 1.207 507 039 343 337 749(55), with a fractional uncertainty of $4.6\times10^{-17}$. Fully benefiting from the clock stability, the ratio measurement provides a powerful probe for new physics by exploring the variation $\Delta\alpha/\alpha$ of the fine structure constant with an uncertainty reducing as $1.6\times10^{-15}\,(\tau/\mathrm{s})^{-1/2}$. This already improves on dedicated experiments using atomic dysprosium[11] in a search for variations on a timescale of seconds[12], motivated by a coupling of light bosonic dark matter[13] to the electromagnetic field.**

While the excitation of an atomic transition is described by a coherent quantum evolution, the readout of the outcome finds atoms in either the ground state or the excited state, introducing quantum projection noise[14] (QPN), which sets a quantum limit on clock stability that scales with atom number as $1/\sqrt{N}$. Consequently, optical lattice clocks operating with $N \gg 1$ offer a substantial advantage over single-ion clocks. Fully exploiting this excellent quantum-limited stability for a stand-alone clock relies crucially on the continued development of ultrastable clock lasers[15], since the noise of the local oscillator in conjunction with the periodic interrogation of the clock transition degrades stability through the Dick effect[16]. This, however, can be rejected in a synchronous comparison of two clocks [17,18], where shared frequency noise of the clock lasers leads to common Dick effect noise in the frequency evaluations, which can be cancelled out in determining the frequency ratio. An extension to this scheme promises interrogation times beyond the coherence time of the lasers[19]. Combining long interrogation times with increased atom numbers, synchronous comparisons of future optical lattice clocks may achieve a statistical uncertainty of $1\times10^{-18}$ within minutes of averaging time[18]. This will allow instant investigation not only of clock frequency ratios, but also of relativistic effects equivalent to elevation changes of a single centimetre.

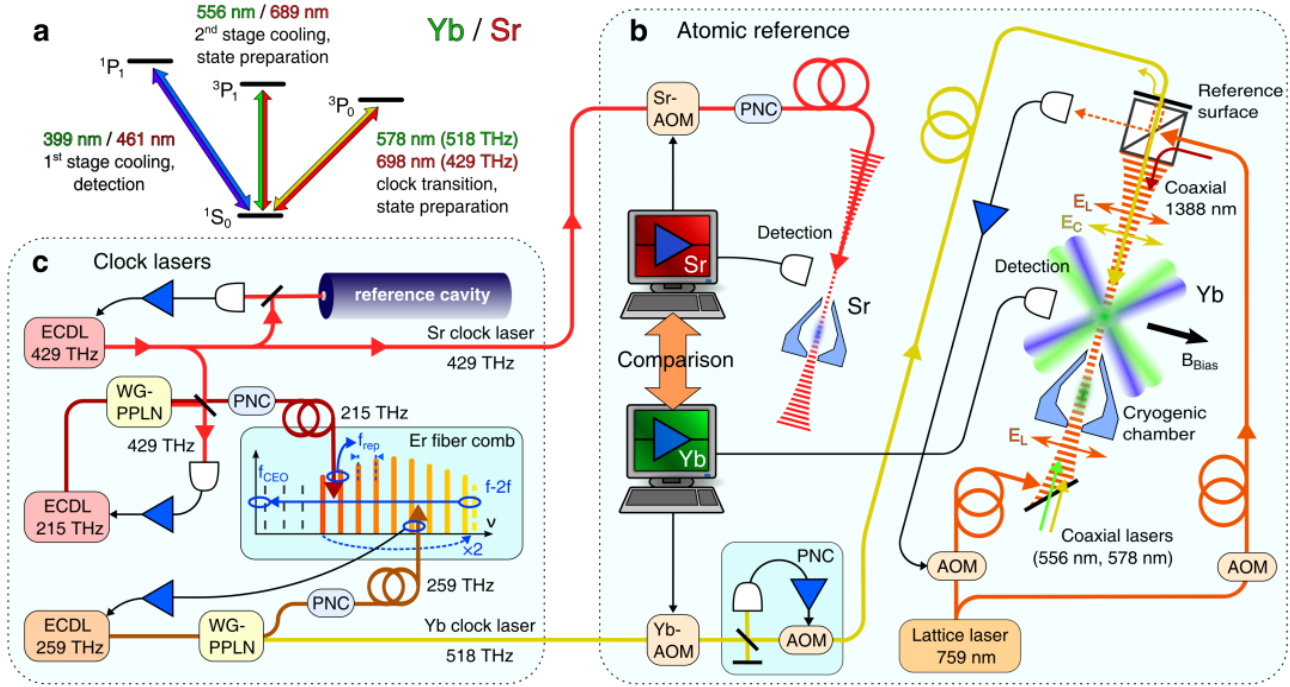

**Figure 1: Experimental setup of Yb/Sr ratio measurement. a,** Relevant transitions in Yb and Sr. **b,** Optical scheme of Yb and Sr clocks. The control systems apply frequency corrections through acousto-optic modulators (AOM). PNC: phase-noise canceller. **c,** Generation of phase-locked clock lasers at $\nu_{Sr}^{cl} \approx 429$ THz and $\nu_{Yb}^{cl} \approx 518$ THz using an Er-fibre comb bridging the frequencies $\nu_{Sr}^{cl} / 2$ and $\nu_{Yb}^{cl} / 2$. Waveguide periodically-poled lithium niobate crystals (WG-PPLN) double the output frequencies of intermediary external-cavity diode lasers (ECDL).

We apply synchronous interrogation in determining the ratio of the clock transition frequencies $\nu_{Yb} \approx 518$ THz and $\nu_{Sr} \approx 429$ THz in neutral $^{171}$Yb and $^{87}$Sr using a pair of cryogenic optical lattice clocks[2], designed for operation with either of the two atomic species. Figure 1 gives an overview of the experimental setup. Laser-cooled atoms loaded into the optical lattice are spin-polarized to more than 95% purity and sideband-cooled (see Methods and Supplementary Fig. 1) to an average axial vibrational state of $\bar{n} < 0.1$. We then probe the clock transitions using phase-locked ultrastable clock lasers.

For clocks interrogating neutral Yb or Sr in a room temperature environment, the largest systematic frequency shift results from the AC Stark effect caused by blackbody radiation (BBR) and significant efforts have been made to control and characterize this effect[20-22]. In our experiments, the atoms are transported to a cryogenic chamber for interrogation. This shields the atoms from ambient BBR and results in a near hundred-fold reduction of the frequency shift; for Yb from −1.278 Hz at 300 K to −14.2(4) mHz at 96 K, and for Sr from −2.278 Hz at 300 K to −23.3(4) mHz at 95 K.

With less than $10^{-18}$ uncertainty of the BBR shifts, the largest uncertainty contribution for the ratio measurement stems from the lattice light shifts in the Yb clock. Taking into account atomic hyperpolarizability[23] and multipolar effects, the light shifts are modelled based on ref. 24 (see Methods and Supplementary Fig. 2 for details). We find $\nu_{E1} = 394{,}798{,}265(9)$ MHz for the E1-magic frequency, where the electric dipole polarizabilities of ground and excited states of the clock transition are equal.

We typically operate the optical lattice at $\nu_{op} = 394{,}798{,}278$ MHz, where the sensitivity to a variation of the lattice intensity is reduced by a partial cancelation of the linear and quadratic terms of the lattice-induced clock frequency shifts. For a typical trap depth $U_0 = 100\ E_r$ ($\approx 10\ \mu$K), where $E_r$ is the lattice photon recoil energy, our light shift model predicts a residual shift of $\Delta\nu_{ls} / \nu_{Yb} = (-12 \pm 32) \times 10^{-18}$. The uncertainty is dominated by the determination of the model parameters, but also accounts for deviations from ideal lattice conditions, as might result from an intensity imbalance or a spectrally broad background in the laser emission (see Methods). The uncertainty budget for the two clocks and the ratio measurement is given in Table 1.

The Sr clock laser is stabilized to a 40-cm-long cavity and shows a stability of 3 to $5 \times 10^{-16}$ at 1 s. Its spectral characteristics are transferred to the Yb clock laser by a tight phase-lock, using an Er-fibre frequency comb with servo bandwidths $\approx 1$ MHz[25]. The control systems for Yb and Sr then independently adjust the laser frequencies through acousto-optic modulators (AOM). After applying an interrogation pulse to the atoms in the optical lattice, the number of atoms in the ground and excited clock states, $N_g$ and $N_e$, are detected to determine the atomic excitation $P = N_e / (N_g + N_e)$, which is

**Table 1: Corrections and uncertainty contributions**

| | $^{171}$Yb | | $^{87}$Sr | |
|---|---|---|---|---|
| **Systematic effect:** | Correction ($10^{-18}$) | Uncertainty ($10^{-18}$) | Correction ($10^{-18}$) | Uncertainty ($10^{-18}$) |
| Quadratic Zeeman effect | 67.7 | 9.8 | 117.0 | 1.0 |
| BBR shift | 27.5 | 0.7 | 54.2 | 0.9 |
| Lattice light shift | 8.5 | 32.8 | 3.5 | 3.4 |
| Probe light shift | -0.8 | 3.2 | 0.09 | 0.05 |
| Collisions | 0.0 | 3.4 | 0.9 | 4.2 |
| AOM chirp and switching | 0.0 | 1.1 | 0.0 | 0.2 |
| 1$^{st}$ order Doppler effect | 0.0 | 2.0 | 0.0 | 0.5 |
| Servo error | 0.8 | 1.1 | 1.9 | 1.6 |
| **Single clock total** | **103.7** | **34.7** | **177.6** | **5.8** |

| **Yb/Sr ratio:** | Correction ($10^{-18}$) | Uncertainty ($10^{-18}$) |
|---|---|---|
| Yb systematic effects | 103.7 | 34.7 |
| Sr systematic effects | -177.6 | 5.8 |
| Laser-to-laser link | 0.0 | 5.4 |
| Gravitational shift | -0.3 | 0.2 |
| Statistical uncertainty | 0.0 | 28.6 |
| **Total** | **-74.2** | **45.6** |

Values are given in fractional units of $10^{-18}$ and represent averages over the complete set of contributing ratio measurements. See ref. 2 and Methods for details. The stated statistical uncertainty represents the standard deviation of the individual ratio measurements. The corrected ratio relates to the uncorrected transition frequencies $\tilde{\nu}_{Yb}$ and $\tilde{\nu}_{Sr}$ as $\mathcal{R} = \tilde{\nu}_{Yb}(1+\Delta_{Yb}) / [\tilde{\nu}_{Sr}(1+\Delta_{Sr})] \approx (1+\Delta_{Yb}-\Delta_{Sr})\,\tilde{\nu}_{Yb}/\tilde{\nu}_{Sr}$, where $\Delta_{Yb}$ and $\Delta_{Sr}$ are the fractional corrections for each transition frequency.

used to estimate the deviation of the laser frequency from the atomic resonance and to calculate the required frequency correction.

We operate both clocks with the same cycle time of 1.5 s and simultaneously apply interrogation pulses of 200 ms duration. As a result, the atomic excitation in both clocks acquires a correlated fluctuation, as indicated in Fig. 2(b). The resulting common-mode noise cancels[18] for the clock frequency ratio $\mathcal{R} = \nu_{Yb} / \nu_{Sr}$, which is directly extracted from the record of applied corrections and resulting atomic excitation.

Figure 2(a) shows the fractional statistical uncertainty of the measured ratio $\mathcal{R}$, which falls as quickly as $4\times10^{-16}\,(\tau/\text{s})^{-1/2}$ and reaches $1\times10^{-17}$ with an averaging time below 3000 s. This is fully competitive with the stabilities of single species comparisons[21,26] that employ state-of-the-art clock lasers with stabilities of $1-2\times10^{-16}$ at 1 s.

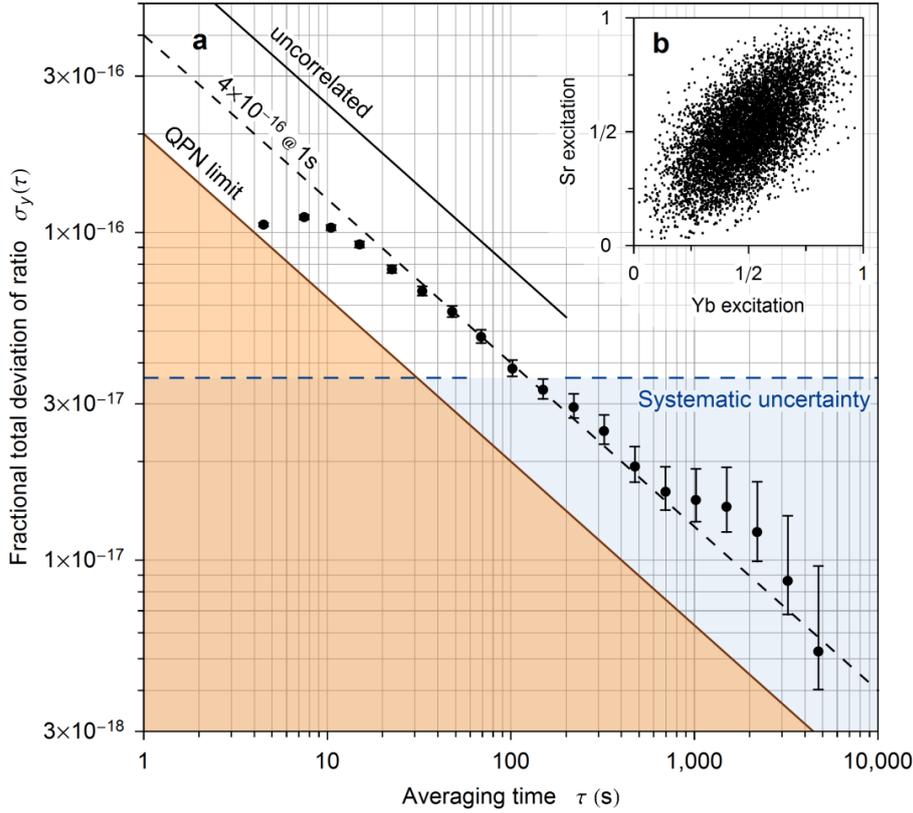

**Figure 2: Ratio measurement stability. a**, The best observed short-term stability (given as fractional total deviation[29]), for a measurement of $10^4$ s improves as $\sigma_y(\tau) = 4\times10^{-16}\,(\tau/\text{s})^{-1/2}$ (black dashed line), approaching the QPN limit of $2\times10^{-16}\,(\tau/\text{s})^{-1/2}$ for $N_{Yb}$ = 500 and $N_{Sr}$ = 1000 (brown line). Error bars indicate $1\sigma$ uncertainties, assuming white frequency noise. The stability estimated for an asynchronous measurement with similar but uncorrelated laser noise improves as $8\times10^{-16}\,(\tau/\text{s})^{-1/2}$ (solid black line). Blue dashed line indicates the systematic ratio uncertainty of $3.6\times10^{-17}$. **b**, Correlation of atomic excitation detected in the Yb and Sr clocks after synchronous application of each clock pulse.

Judging from the evaluation of interleaved Yb measurements, we find the synchronous interrogation to yield a two times improvement in stability compared to a measurement with uncorrelated laser noise of similar magnitude (see Methods). The results demonstrate the excess noise introduced by the comb-based phase-lock to be low enough to approach the QPN limit of $2\times10^{-16}\,(\tau/\text{s})^{-1/2}$ for 500 Yb and 1000 Sr atoms to a factor of two. For $\tau = 150$ s, the stability has already surpassed the systematic uncertainty of $3.6\times10^{-17}$, such that the overall uncertainty of the ratio measurement reaches $4.8\times10^{-17}$.

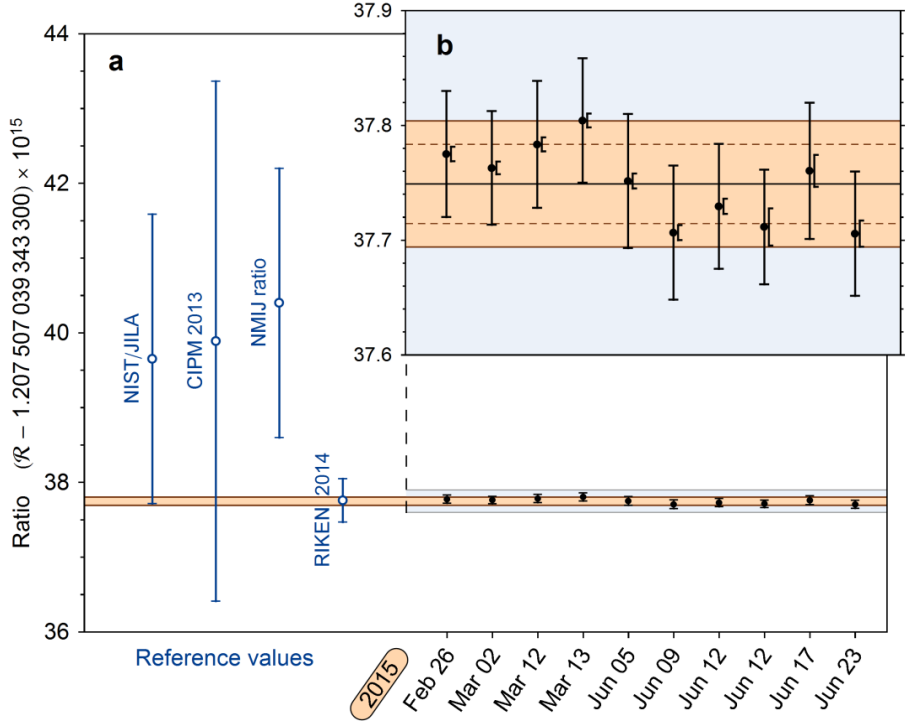

**Figure 3: History of ratio measurements. a,** Orange shaded region marks the 1σ uncertainty range of $\mathcal{R}$. Black circles show results of individual measurements, with error bars indicating total uncertainties. Ratios determined from previous measurements are shown in blue: "CIPM 2013" is calculated from the CIPM recommended frequencies[28], "NIST/JILA" from the frequency measurements with the lowest uncertainties listed therein, performed against the caesium standard at NIST. "NMIJ ratio" results from a direct ratio measurement[7]. "RIKEN 2014" represents a measurement prior to those presented here[27]. **b,** Enlargement of blue-shaded region. Additional smaller error bars indicate statistical uncertainties of individual measurements, dashed lines show their standard deviation.

Over a period of four months, we performed ten measurements that are used to determine the Yb/Sr frequency ratio (Fig. 3). Due to the good short-term stability, the reproducibility of the measurements is not limited by the available averaging time, but suggests a variation of uncontrolled systematic effects between repeated experiments. We therefore determine a statistical uncertainty for the entire series of measurements from the standard deviation of $\sigma_\mathcal{R}/\mathcal{R} = 2.9 \times 10^{-17}$, with no reduction according to the standard error of the mean. With equal weight assigned to each measurement, we find $\mathcal{R} = \nu_{Yb}/\nu_{Sr} = 1.207\,507\,039\,343\,337\,749(43)_{sys}(35)_{stat}$, consistent with our prior measurement[27]. The fractional uncertainty of $4.6 \times 10^{-17}$ represents a 30-fold reduction compared to a previous direct measurement at NMIJ[7], which yielded $\mathcal{R}^{NMIJ} = 1.207\,507\,039\,343\,340\,4(18)$. Our value deviates from $\mathcal{R}^{NMIJ}$ by 1.5 times their stated uncertainty, but is well within the uncertainty of the ratio $\mathcal{R}^{\text{CIPM}} = 1.207\ 507\ 039\ 343\ 339\ 9(35)$ calculated from the transition frequencies recommended by the International Committee for Weights and Measures (CIPM)[28] in 2013.

Our ratio measurement obtains an overall uncertainty that improves on the best reported value. With enhanced control of the systematic effects, the measurement stability promises a statistical uncertainty of $1\times10^{-18}$ with less than two days of averaging time. Operation with larger atom numbers will further relax the QPN limit and -together with improvements to clock laser stability- allow additional reductions in averaging times, making ratio measurements between optical lattice clocks a valuable tool in a search for new physics[11-13] at short time scales.

## Methods

**Operation of the Yb clock.** Fermionic $^{171}$Yb ($I=1/2$) atoms are decelerated from a thermal beam that contains both Sr and Yb atoms by a Zeeman slower and trapped in a magneto-optical trap (MOT), operating on the $^1S_0 - {}^1P_1$ transition at 399 nm. A second cooling stage in a MOT on the $^1S_0 - {}^3P_1$ transition at 556 nm reduces the atomic temperature to 15 µK. The optical lattice is kept at full intensity during loading. It provides a trap depth $U_0 / k_B \approx 10$ µK or $U_0 \approx 100\ E_r$, with the lattice photon recoil energy $E_r = (h / \lambda_L)^2 / (2m)$, where $\lambda_L \approx 759$ nm is the lattice laser wavelength, and $m$ is the atomic mass. Around 500 atoms at a temperature of 3 µK remain in the lattice after turning off the MOT beams. Changing the frequency of one of the two independent lattice beams creates a moving lattice, which transports the atoms by a chosen distance of typically 18−20 mm into the cryogenic chamber located below the MOT region (see Fig. 1). Outside of the 60 ms transport interval, the relative phase of the lattice beams is actively stabilized to fix the lattice anti-nodes relative to a reference mirror[2].

To interact with the atoms inside the cryogenic chamber, additional coaxial laser beams (see Fig. 1b) are superimposed on the optical lattice with the same linear polarization. Spin-polarization is realized by optical pumping on the $^1S_0$ ($F=1/2$) − $^3P_1$ ($F=3/2$) transition, where a bias magnetic field of 70 µT separates the Zeeman components by 900 kHz and allows the $m_F = \pm1/2$ to $m_{F'} = \pm1/2$ π−transitions to be independently addressed. Two additional beams are used for axial sideband cooling: The first beam excites the red sideband of the $^1S_0 - {}^3P_0$ clock transition at 578 nm with a Rabi frequency of $\Omega_R^{\text{red}} > 2\pi\times100$ Hz. The second beam is resonant with the $^3P_0$ ($F=1/2$) − $^3D_1$ ($F=1/2$) transition at 1388 nm and returns the atoms to the ground state. Simultaneous operation of all three beams accumulates spin-polarized atoms in the dark $n = 0$ axial vibrational state over a period of 100 ms (see Supplementary Fig. 1). We typically find >95% of atoms in the desired spin state and an average axial vibrational quantum number $\bar{n} \approx 0.08$.

A Rabi interrogation pulse of 200 ms is applied using a separate 578 nm beam (see Fig. 1b), which is phase-stabilized[30] to the same reference surface as the lattice. After interrogation, the atoms are transported back to the loading region by the moving lattice and a sequence of three camera images is acquired to extract the atomic excitation from the background-corrected fluorescence signals of atoms in the ground and excited state. Over a 4-step sequence, the clock laser probes the left and right slopes of the $m_F = \pm 1/2$ components to measure and correct for the clock laser drift while cancelling out the first order Zeeman shift and the vector Stark shift[31].

**Ratio determination from recorded steering data.** The control systems store the applied AOM frequencies $\Delta f_{Yb(Sr)}^{AO}$ and resulting atomic excitation $P$. The excitation is used to calculate the error signal $\delta v$, corresponding to the deviation of the applied clock laser frequency from the HWHM point of the Rabi lineshape.

The subharmonics of the clock laser frequencies, $v_{Sr}^{cl}/2$ and $v_{Yb}^{cl}/2$, are used to create beat signals with the comb lines with indices $n_{Sr}$ and $n_{Yb}$. The clock frequency ratio $\mathcal{R} = v_{Yb}/v_{Sr}$ can be extracted from recorded data as

$$\mathcal{R} = r + \frac{(f_{CEO} + \Delta f_{Yb}^*/2) - r(f_{CEO} + \Delta f_{Sr}^*/2)}{n_{Sr} f_{rep} + (f_{CEO} + \Delta f_{Sr}^*/2)}, \qquad (1)$$

where $r = n_{Yb}/n_{Sr}$, and $f_{rep}$ and $f_{CEO}$ are the comb repetition rate and carrier-envelope offset frequency. The frequency offsets $\Delta f_{Yb(Sr)}^* = \Delta f_{Yb(Sr)}^{AO} - \Delta v_{Yb(Sr)}^{step} + \Delta f_{Yb(Sr)}^{const} + \Delta v_{Yb(Sr)}^{cor} - \delta v_{Yb(Sr)}$ include the AOM frequency $\Delta f_{Yb(Sr)}^{AO}$, corrected for the frequency step $\Delta v_{Yb(Sr)}^{step}$ applied to probe the left and right slopes of the two Zeeman components. All constant frequency offsets are included in $\Delta f_{Yb(Sr)}^{const}$, while $\Delta v_{Yb(Sr)}^{cor}$ corrects for systematic shifts. The division by two accounts for the use of the subharmonics. To avoid counting errors from a direct measurement of $f_{rep}$, we substitute $n_{Sr} f_{rep} + (f_{CEO} + \Delta f_{Sr}^*/2) = v_{Sr}/2$, which is known to a fractional uncertainty of less than $10^{-15}$. All radio-frequencies (RF) are derived from a GPS-conditioned BVA quartz oscillator with a stability of $\leq 5 \times 10^{-12}$ for $\tau > 0.2$ s. With proper choice of the referenced comb lines, the last term in eq. (1) is on the order of $10^{-7}$ and the RF stability contributes an uncertainty $<10^{-18}$, as discussed later.

**Short-term stability during Yb/Sr comparisons and interleaved evaluation.** In a simple model of the Dick effect contribution to the short-term stability based on ref. 32, we treat the Sr clock laser as dominated by cavity thermal noise, with a constant flicker FM instability of $\sigma_y(\tau) \approx 3 \times 10^{-16}$, and the link between clock lasers as a source of white phase noise chosen to reproduce $\sigma_y^{sync}(\tau) = 4 \times 10^{-16} (\tau/s)^{-1/2}$. The model predicts a stability of $1.4 \times 10^{-15} (\tau/s)^{-1/2}$ for an interleaved measurement in the Yb clock with a cycle time of 3 s, in agreement with experimental results under optimal conditions. For a comparison of Yb and Sr clocks with uncorrelated noise, the predicted stability for a cycle time of 1.5 s is $\sigma_y^{\mathrm{async}}(\tau) = 8\times10^{-16}\,(\tau/\mathrm{s})^{-1/2}$, with contributions from individual clock stabilities of $\sigma_y^{\mathrm{Sr}}(\tau) = 5\times10^{-16}\,(\tau/\mathrm{s})^{-1/2}$ and $\sigma_y^{\mathrm{Yb}}(\tau) = 6\times10^{-16}\,(\tau/\mathrm{s})^{-1/2}$.

**Sensitivity to changes in the fine structure constant.** To the first order, the measured frequency ratio varies with $\alpha$ as $\Delta\mathcal{R}/\mathcal{R} = (A_{\mathrm{Yb}} - A_{\mathrm{Sr}})\,\Delta\alpha/\alpha$, where $A_{\mathrm{Yb}} = 0.31$ and $A_{\mathrm{Sr}} = 0.06$ are sensitivity factors for Yb and Sr[33,34]. Certain ion clocks, such as those using Hg$^+$ ($A_{\mathrm{Hg+}} = -3.2$) or the octupole clock transition in Yb$^+$ ($A_{\mathrm{Yb+}}^{\mathrm{oct}} = -5.3$) exhibit larger sensitivities and offer a clear advantage in a search for a long term drift $\dot{\alpha}$ over a period of months or years[35]. When looking for a sudden change $\Delta\alpha$ associated with the passage of topological defect dark matter[12], the longer averaging times required by ion clocks negate this advantage at least partially: For the Al$^+$/Hg$^+$ ratio[5], a stability of $3.9\times10^{-15}\,(\tau/\mathrm{s})^{-1/2}$ was reported, equivalent to a sensitivity to $\Delta\alpha/\alpha$ improving as $\sigma_y^{\alpha}(\tau) = 1.2\times10^{-15}\,(\tau/\mathrm{s})^{-1/2}$. For a Yb$^+$/Sr comparison[36] the reported stability was $5.3\times10^{-15}\,(\tau/\mathrm{s})^{-1/2}$, corresponding to $\sigma_y^{\alpha}(\tau) = 1.0\times10^{-15}\,(\tau/\mathrm{s})^{-1/2}$. In terms of the limit that can be set on $\Delta\alpha/\alpha$ for short averaging times, the sensitivity of $\sigma_y^{\alpha}(\tau) = 1.6\times10^{-15}\,(\tau/\mathrm{s})^{-1/2}$ for our Yb/Sr measurement is therefore already competitive with existing experiments. Further improvements are expected with increased atom number.

**Uncertainty budget of the Yb clock.** The following sections discuss the uncertainty contributions stated in Table 1 for the Yb clock.

**Quadratic Zeeman shift.** The alternating interrogation of the $m_F = \pm 1/2$ components provides a continuous measurement of the magnetic field. We use the coefficients reported in ref. 37 to correct for the resulting quadratic Zeeman shift.

**BBR shift.** A detailed study of the residual BBR frequency shifts for atoms inside the cryogenic chamber and their uncertainty was performed for Sr (ref. 2). As the contribution from the dynamic polarizability is insignificant at $T = 96$ K, the twice lower DC-polarizability of Yb compared to Sr leads to a further reduction of shift and uncertainty. Precise experimental polarizability coefficients are available in refs. 22,38.

As the lattice waist is located between the loading position and the entrance aperture of the cryogenic chamber, the trap depth falls with increased transport distance inside the chamber. To avoid excessive atom loss in the Yb clock, where the temperature of the trapped atoms is higher than for Sr, atoms are interrogated at a chosen position located 5-7 mm inside the chamber, as opposed to 9 mm for Sr. The increased effect of room-temperature BBR leaking into the chamber is included in the uncertainty contribution of $7\times10^{-19}$.

**Lattice light shifts.** In Yb, the presence of several allowed two-photon transitions close to the magic frequency introduces hyperpolarizability effects large enough to require consideration even at the $10^{-17}$ level[23]. The magnitude of the multipolar polarizability has also been studied only theoretically. To include both effects in our evaluation, we model the lattice-induced light shift $\Delta\nu_{ls}$ as

$$\Delta\nu_{ls} = (a\,\Delta\nu_L - b)\left(\bar{n} + \tfrac{1}{2}\right)\sqrt{U_e/E_r} - \left(a\,\Delta\nu_L + \tfrac{3}{4}d\,(2\bar{n}^2 + 2\bar{n} + 1)\right)U_e/E_r$$
$$+ d(2\bar{n} + 1)(U_e/E_r)^{3/2} - d(U_e/E_r)^2\,, \tag{2}$$

derived from eq. (12) of ref. 24 by substituting the single beam intensity $I = U_e/\alpha_{E1}$, where the atomic electric dipole (E1) polarizability $\alpha_{E1}$ is identical for ground and excited clock states at the E1-magic frequency $\nu_{E1}$. The equivalent axial trap depth $U_e = \zeta U_0$ represents the reduced intensity experienced by the atoms due to their radial motion[8]. The coefficient $a$ describes the slope of the differential E1-polarizability at $\nu_{E1}$, while the coefficients $b$ and $d$ correspond to the combined multipolar polarizability and hyperpolarizability, respectively. $\Delta\nu_L = \nu_L - \nu_{E1}$ is the detuning of the lattice frequency $\nu_L$ from $\nu_{E1}$. Replacing the individual atomic axial vibrational quantum number $n$ by the ensemble average $\bar{n}$ is a convenient approximation for our experiments, where $\bar{n} < 1$. The parameters $U_0$, $\zeta$ and $\bar{n}$ are extracted from axial sideband spectra for each measurement using methods based on ref. 39. For the hyperpolarizability coefficient we adopt a value of $d = -1.9(8)$ µHz to reproduce the results of ref. 23. The increased uncertainty results mainly from our estimation of $\zeta_{NIST} = 0.70(14)$.

To find $a$, $b$ and $\nu_{E1}$, the light shift model is simultaneously fitted to the data accumulated in 50 interleaved measurements, varying either the trap depth or –by alternating between operation with and without sideband cooling– the vibrational state (see Supplementary Fig. 1 and 2). When determining the weighting for this fit, we consider the uncertainties of the determined parameters, particularly the equivalent trap depth $U_e$, in addition to the statistical uncertainty of the measurement itself. This avoids introducing errors by putting excessive weight on individual measurements that depend critically on these input parameters, such as light shift measurements far detuned from the magic frequency. All statistical uncertainties are inflated by $\sqrt{\chi^2} = 1.2$ based on a subset of 29 measurements within 100 MHz of the magic frequency, where the parameter dependence is small compared to the statistical uncertainty of a single measurement.

The uncertainties of the coefficients extracted from the fit represent uncertainties uncorrelated between measurements. To take into account systematic errors that are shared between measurements, specifically in our determination of $\zeta$ and $\bar{n}$ from the sideband spectra and the adopted value of $d$, we employ a Monte-Carlo method for our data analysis. In this procedure, fitting is repeated for different permutations of these parameters and the RMS deviation of the obtained coefficients is added to their uncertainties in quadrature. The effects of collisional shifts on the light shift evaluation are also included at this stage.

Overall, we find $a = 0.021(6)$ mHz/MHz and $b = -0.68(71)$ mHz, as well as $\nu_{E1} = 394{,}798{,}265(9)$ MHz. Typical parameter values are $\zeta = 0.72(5)$ and $\bar{n} = 0.08(8)$ when sideband cooling is applied.

**Lattice light impurity.** The optical lattice is generated by the output of a Ti:Sapphire laser, filtered with a volume Bragg grating (VBG) with a FWHM of 40 GHz. The lattice frequency is determined to within 200 kHz using the frequency comb. The optical spectrum shows no amplified spontaneous emission above the noise floor of the spectrum analyser, and a monitoring Fabry-Pérot etalon shows no spurious frequency components during normal operation.

Although we do not expect a frequency shift, we perform a check by deliberately detuning the centre frequency of the VBG by $f_{\text{rel}}^{\text{VBG}} = \pm 10$ GHz from the lattice laser frequency $\nu_L$ to impose an asymmetric shape on a possible spectrally broad background. We then measure the light shift by varying the lattice intensity. Extrapolated to a trap depth of $U_0 = 100\ E_r$, the results correspond to a fractional frequency shift of $(\Delta\nu_{\text{bg}}/\nu_{\text{Yb}})/\Delta f_{\text{rel}}^{\text{VBG}} = -1.9(1.4)\times10^{-18}$/GHz. For an alignment of the VBG centre-frequency according to $f_{\text{rel}}^{\text{VBG}} = 0(5)$ GHz in normal operation, we include a fractional uncertainty of $1.2\times10^{-17}$ (at 100 $E_r$) in the lattice light shift uncertainty.

**Running wave frequency shifts.** The multipolar polarizability assumed in our light shift model contributes a frequency shift of $\Delta\nu_{\text{mp}}/\nu_{\text{Yb}} \approx -b(\bar{n} + 1/2)\sqrt{U_e/E_r} = 6.4(6.8)\times10^{-18}$ for nominal trapping parameters (eq. (2)). This shift varies with the shape of the trapping potential, with the model assuming a pure standing-wave field. An imbalance in lattice beam intensity creates an admixture of running-wave field that may cause a frequency shift[40] $\Delta\nu_{\text{rw}}$.

To investigate the effect, a measurement is performed near the beam waist, where a trap depth of $U_0 = 165\ E_r$ at full lattice intensity corresponds to a single beam peak intensity of $I = 10$ kW/cm$^2$. Reducing the intensity of one lattice beam to $I_1 = 1.5$ kW/cm$^2$ results in a strong intensity imbalance $\Delta I = I_2 - I_1 = 8.5$ kW/cm$^2$. Interleaved measurements alternate between this imbalanced configuration and a balanced configuration with $I_1 = I_2 = 6$ kW/cm$^2$, corresponding to the intensity inside the cryogenic chamber. We find a statistically insignificant frequency shift of $\Delta\nu_{\text{rw}}/\nu_{\text{Yb}} = 1.4(2.8)\times10^{-17}$. For a linear model $\Delta\nu_{\text{rw}}/\nu_{\text{Yb}} = c_{\text{rw}}|\Delta I|$ with $c_{\text{rw}} = 1.6(3.3)\times10^{-18}$/(kW/cm$^2$) the maximum expected imbalance of $\Delta I = 1.2$ kW/cm$^2$ then yields a frequency shift of $2.0(3.9)\times10^{-18}$. To avoid introducing an unwarranted frequency correction, we include this in the lattice light shift uncertainty budget as $0.0(5.9)\times10^{-18}$.

**Probe light shift.** The intensity required to apply a Rabi $\pi$-pulse over the interrogation time $t_i$ falls as $I_{\text{pr}} \propto t_i^{-2}$, strongly reducing frequency shifts for longer interrogation times. We use the values reported in ref. 37, rescaled to $t_i = 200$ ms, to determine the probe light shift and its uncertainty.

**Collisional shifts.** For a typical atom number of $N_{Yb} = 500$, the average population per lattice site is less than one. We expect a suppression of *p*-wave collisions due to the low atomic temperature of 3 μK and trap depth equivalent to 10 μK, both significantly below the *p*-wave barrier of ≥ 30 μK[41]. In a spin-polarized and homogeneously excited sample, the Pauli exclusion principle prohibits *s*-wave collisions.

We investigate density-induced frequency shifts by measurements alternating between high and low atom numbers and observe no statistically significant effect over repeated measurements. For the lattice configuration inside the cryogenic chamber, a trap depth of $U_0 = 100\ E_r$ and $N_{Yb} = 500$, we adopt a fractional uncertainty of $4\times10^{-18}$. The Monte-Carlo stage of the light shift analysis also varies the collisional shift coefficient according to this uncertainty. To include the effect of density changes with trap depth, we apply a scaling according to $U_0^{3/2}$ (ref. 3).

**First-order Doppler shift.** The phase of the clock laser and the positions of the lattice antinodes are actively stabilised to the same reference surface[2]. We include an uncertainty of $0.5\times10^{-18}$ to account for residual variations in the clock laser path length.

In our experiments, the lattice axis is oriented at a 15° angle from vertical, such that gravity causes a transverse displacement of the atomic distribution that varies with radial confinement. Changes in lattice intensity can therefore induce atomic motion during interrogation. This motion follows the wavefront of the lattice, and unless the wave-vector of the clock laser is perfectly perpendicular, will result in an additional Doppler shift. We investigate the effect by applying lattice intensity ramps characterized by $\dot{U}_0 = dU_0/dt \approx \pm 125\ E_r/s$ during interrogation. The largest observed frequency shifts are equivalent to $(\Delta\nu_{Dop}/\nu_{Yb}) / \dot{U}_0 \approx 2\times10^{-18} / (E_r/s)$.

The intensities of the lattice beams are not actively stabilized and persistent intensity changes correlated with the experimental cycle might be introduced by thermal effects in the lattice AOMs. However, even following an intensity reduction from 100% to 85% during the state preparation period, we observe intensity changes corresponding to only $\dot{U}_0 = 0.0(1.0)\ E_r/s$. We adopt an uncertainty of $1.9\times10^{-18}$ due to induced transverse motion, for a total first-order Doppler-shift uncertainty of $2\times10^{-18}$.

**AOM chirp and switching.** The AOM used to apply frequency corrections to the clock laser is operated at an RF power of only 20 mW to reduce thermal effects. Instead of being switched off outside the interrogation time, the applied frequency is simply detuned by 100 kHz. At the beginning of the clock interrogation, the frequency switch to resonance causes a reproducible phase excursion of the phase-noise cancellation system. We calculate a resulting frequency shift of $3\times10^{-19}$ (ref. 42). This is further suppressed by randomly alternating between positive and negative detuning, for which the phase excursion likewise changes in sign. A loop-back test confirms the absence of phase drift and chirp: An additional optical fibre returns the light sent to the experimental chamber back to the clock laser, where a partial mirror serves as phase reference. The beat with the unshifted clock laser light directly after the WG-PPLN (see Fig. 1) is then measured by a frequency counter gated to the interrogation time. We find no frequency deviation or instability larger than $1.1\times10^{-18}$, which we take as the uncertainty.

**Servo error.** Both control systems contain algorithms to determine and predict the drift of the clock lasers. For a combined dataset of all ratio measurements, we find that the mean in-loop error signal $\overline{\delta\nu}$, extracted from the detected atomic excitation, reaches $-0.8(1.1)\times10^{-18}$ for Yb and $-1.9(1.6)\times10^{-18}$ for Sr, where the instabilities have been determined from the Allan deviation.

The effect of the servo error on the measured frequency ratio is corrected shot-to-shot by including $\delta\nu$ in the $\Delta f_{Yb}^*$ and $\Delta f_{Sr}^*$ terms of eq. (1). Based on the instability of $\overline{\delta\nu}$, we adopt uncertainty contributions from a long-term servo error of $\sigma_{sv}^{Yb} = 1.1\times10^{-18}$ and $\sigma_{sv}^{Sr} = 1.6\times10^{-18}$.

**DC Stark shifts.** During interrogation, the atoms are shielded from external electric fields by the copper construction of the cryogenic chamber. Although the inner surfaces are black-coated to suppress reflections of room-temperature BBR, a surface resistivity of $R_s < 2\times10^6\,\Omega\,/\,\square$ avoids the build-up of stray charges. We therefore do not expect any significant DC Stark shift. This is supported by the results of experiments operating both clocks with Sr, which find agreement throughout the period spanned by the ratio measurements.

**Additional uncertainties for ratio measurements.** The following sections describe additional uncertainties in the comparison of the two clocks.

**Laser-to-laser link uncertainty.** The frequency comb was constructed at AIST/NMIJ, where the frequency stability of a near-identical comb was evaluated[43] as approximately $2.3\times10^{-18}$ for $\tau = 10^3$ to $10^4$ s and falling to below $10^{-18}$ for $\tau > 10^4$ s. Since we operate the comb in an environment temperature-controlled to 0.1 K, we expect similar stability.

Phase instabilities from variations in the optical path lengths between the clock lasers and the comb also contribute here. To avoid including the unstabilized lengths of the in-coupling fibres of the WG-PPLN systems, the stabilization of the comb repetition rate and the Yb clock laser use the residual fundamental light transmitted through the crystal, where the frequency doubling process ensures a stable phase relation with the generated harmonic output (Fig. 1). The fibres connecting the clock lasers to the comb are equipped with phase-noise cancellation systems[30]. Based on the stability of the loop-back measurements described above, which include a large fraction of the residual unstabilized optical path (approximately 2 m combined between Yb and Sr systems), we adopt an uncertainty of $2\times10^{-18}$.

The stability of the RF reference affects the ratio measurements through the last term of eq. (1). For the first four measurements we include an uncertainty contribution of $7\times10^{-18}$. The applied frequency shifts were then modified to reduce this uncertainty to $3\times10^{-19}$ for measurements from June onwards. The average uncertainty due to RF instability is $4.4\times10^{-18}$ over all evaluated measurements, for a total fractional uncertainty of $5.4\times10^{-18}$ resulting from the laser-to-laser link.

**Gravitational frequency shift.** Both clocks share the same construction and are installed on the same optical table. The stated gravitational shift correction of $-0.3(2)\times10^{-18}$ accounts for a transport distance that is 2−4 mm less for Yb than for Sr, with a positioning uncertainty of 1 mm for each clock.

## Acknowledgments

We thank H. Inaba and F.-L. Hong of AIST/NMIJ for providing the frequency comb. We thank T. Pruttivarasin for commenting on the manuscript. This work received partial support from the Japan Society for the Promotion of Science (JSPS) through its Funding Program for World-Leading Innovative R&D on Science and Technology (FIRST) and from the Photon Frontier Network Program of the Ministry of Education, Culture, Sports, Science and Technology (MEXT), Japan. N.N. acknowledges RIKEN's Foreign Postdoctoral Researcher (FPR) program for financial support.

## Author Contributions

H.K. initiated and coordinated the experiments. M.T. and I.U. characterized and operated the Sr clock, N.N. and T.O. the Yb clock. N.O. maintained and operated the frequency comb. All authors contributed to the experimental setups, discussed the results and commented on the manuscript.

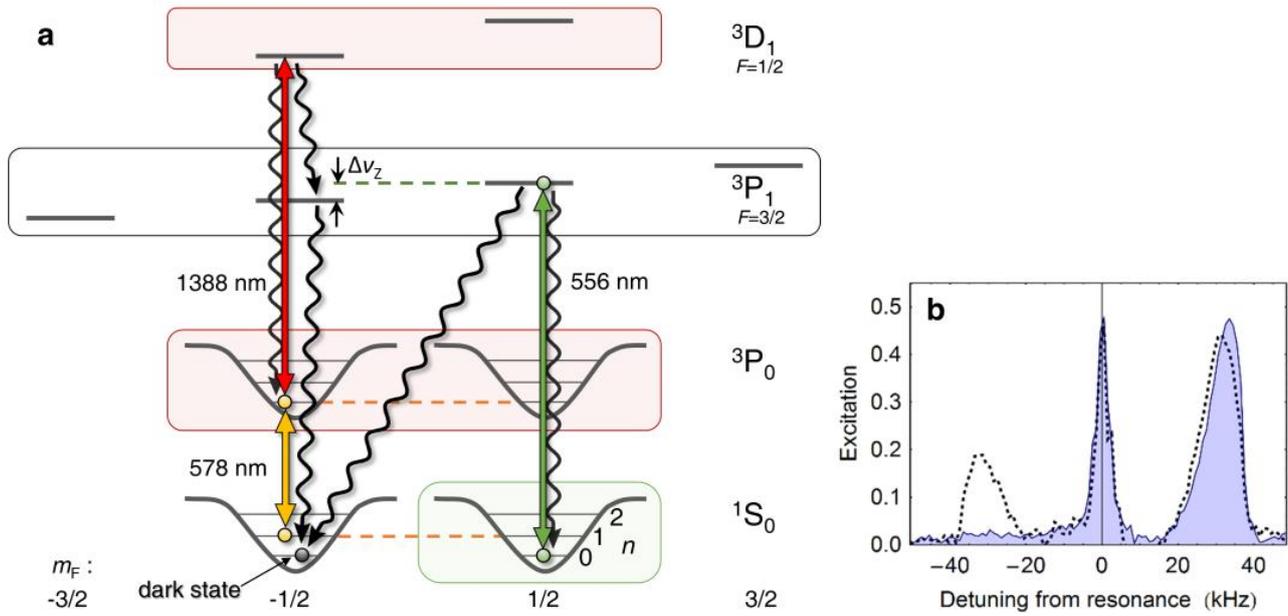

**Supplementary Figure 1: Yb state preparation. a**, Relevant excitation and decay paths. Coloured areas indicate manifolds of substates addressed by the respective laser. Atoms are excited from an axial vibrational state $n$ in $^1S_0$ ($m_F = \pm 1/2$) to $n-1$ in $^3P_0$ ($m_F = \pm 1/2$) by a laser tuned near the edge of the red sideband of the clock transition at 578 nm. The resulting population in the $^3P_0$ state is excited to the short-lived $^3D_1$ ($F=1/2$) state, which provides a decay path to the ground state via $^3P_1$. While the vibrational state is not resolved for the $^1S_0$ to $^3P_1$ transition, the Zeeman splitting $\Delta\nu_Z$ of the $^3P_1$ ($F=3/2$) state allows for spin-polarization to $m_F = -1/2$ [$m_F = 1/2$] by exciting the $m_F = 1/2$ to $m_{F'} = 1/2$ [$m_F = -1/2$ to $m_{F'} = -1/2$] component. While the atoms may return to the $m_F = 1/2$ substate or to a vibrational state $n>0$ via alternative decay paths (not shown), they gradually accumulate in the dark state with $n = 0$ in $^1S_0$ ($m_F = -1/2$). **b,** Sideband spectrum showing a reduction of the average axial vibrational state from $\bar{n} = 0.7$ without sideband cooling (dashed line) to $\bar{n} \approx 0.04$ with sideband cooling (solid, blue shaded curve).

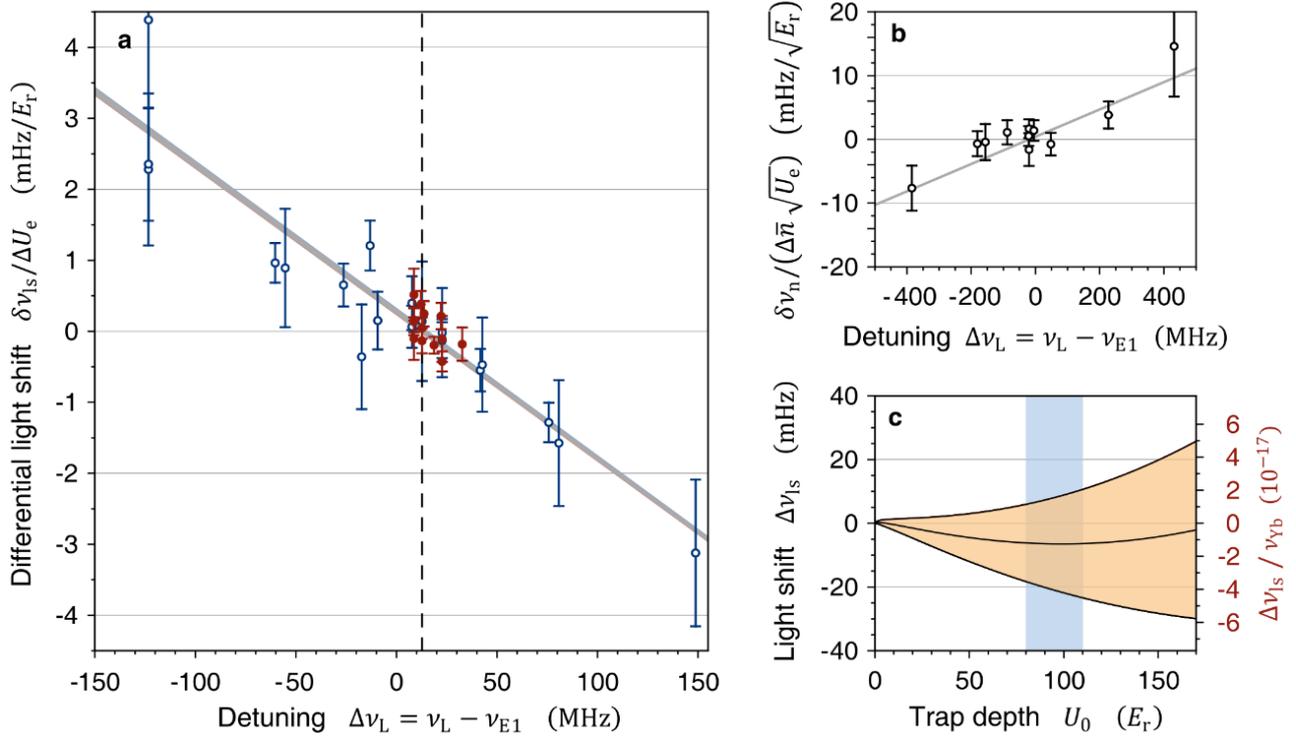

**Supplementary Figure 2: Characterization of lattice light shifts. a,** Measured differential frequency shift $\delta\nu_{ls}$ with change of equivalent trap depth $U_e = \zeta U_0$ as function of lattice detuning $\Delta\nu_L$ from $\nu_{E1}$ = 394,798,265 MHz. Blue open circles indicate measurements alternating between $U_0^{hi} \approx 100\ E_r$ and $U_0^{lo} \approx 70\ E_r$ inside the cryogenic chamber. Red full circles indicate measurements outside the chamber, where proximity to the beam waist allows alternation between $U_0^{hi} \approx 160\ E_r$ and $U_0^{lo} \approx 90\ E_r$. Error bars show 1σ statistical uncertainties after inflation. The grey line indicates the prediction of the light shift model, its width covers typical parameters for both types of measurements. Additional measurements (not shown) were taken at $|\Delta\nu_L| \geq 1$ GHz to accurately determine the detuning dependence of the light shift. Vertical dashed line marks the operating point $\nu_{op}$. **b,** Frequency shift with change of average axial vibrational state $\bar{n}$, plotted as $\delta\nu_n / (\Delta\bar{n}\ U_e^{1/2})$ for normalization. Error bars give inflated statistical uncertainties and the grey line shows the prediction for a measurement alternating between typical values of $\bar{n}^{hi}$ = 0.8 and $\bar{n}^{lo}$ = 0.08 for $U_0$ = 100 to 160 $E_r$. **c,** Estimated light shift and uncertainty of the model as function of trap depth $U_0$ for $\nu_L = \nu_{op}$ = 394,798,278 MHz, blue shaded region indicates the range of $U_0$ typical for our experiments.